\documentclass[conference]{IEEEtran}
\IEEEoverridecommandlockouts
\usepackage{cite}
\usepackage{amsmath,amssymb,amsfonts}
\usepackage{algorithmic}
\usepackage{graphicx}
\usepackage{array}
\usepackage{textcomp}
\usepackage{xcolor}
\def\BibTeX{{\rm B\kern-.05em{\sc i\kern-.025em b}\kern-.08em
    T\kern-.1667em\lower.7ex\hbox{E}\kern-.125emX}}
    
\usepackage{todonotes}

\begin{document}

\title{Detection, Explanation and Filtering of Cyber Attacks Combining Symbolic and Sub-Symbolic Methods}

\author{\IEEEauthorblockN{ Anna Himmelhuber$^1$, Dominik Dold$^2$, Stephan Grimm$^1$,  Sonja Zillner$^1$, Thomas Runkler$^1$}
\IEEEauthorblockA{\textit{$^1$ Siemens AG} \\
Munich, Germany \\
\{anna.himmelhuber, stephan.grimm, sonja.zillner, thomas.runkler\}@siemens.com}
\IEEEauthorblockA{\textit{$^2$ European Space Agency} \\
Noordwijk, Netherlands \\
\{dominik.dold\}@esa.int}
}

\maketitle

\begin{abstract}
Machine learning (ML) on graph-structured data has recently received deepened interest in the context of intrusion detection in the cybersecurity domain. Due to the increasing amounts of data generated by monitoring tools as well as more and more sophisticated attacks, these ML methods are gaining traction. Knowledge
graphs and their corresponding learning techniques such as Graph Neural Networks (GNNs) with their ability to seamlessly integrate data from multiple domains using human-understandable vocabularies, are finding application in the cybersecurity domain. However, similar to other connectionist models, GNNs are lacking transparency in their decision making. This is especially important as there tend to be a high number of false positive alerts in the cybersecurity domain, such that triage needs to be done by domain experts, requiring a lot of man power. Therefore,  we are addressing Explainable AI (XAI) for GNNs to enhance trust management by exploring combining symbolic and sub-symbolic methods in the area of cybersecurity that incorporate domain knowledge. We experimented with this
approach by generating explanations in an industrial demonstrator system.  The proposed method is shown to produce intuitive explanations for alerts for a diverse range of scenarios. Not only do the explanations provide deeper insights into the alerts, but they also lead to a reduction of false positive alerts by 66\% and by 93\% when including the fidelity metric. 

\end{abstract}


\section{Introduction}
The continuous increase in cyber attacks has given rise to a growing demand for modern intrusion detection approaches that leverage ML to detect both simple security risks as well as sophisticated cyber attacks \cite{b1}.  These approaches identify patterns in data and highlight anomalies corresponding to attacks. Such detection tasks are particularly poised to benefit from the ability to automatically analyze and learn from vast quantities of data. There are many relevant examples of the application of deep learning and similar techniques for intrusion detection systems (IDS) \cite{b14} based on anomaly detection methods able to find deviations from a previously learned baseline \cite{b3}. However, their drawbacks include alarm flooding problems \cite{b13} and a lack of explainability, e.g., for why certain network traffic is flagged  as anomalous by the IDS \cite{b2}.\\
This is not just relevant for the defense of conventional IT systems, but also in the context of modern operational technology (OT) systems, such as those used in factories and other industrial automation settings. While these industrial control systems used to be exclusively deterministic in their operation, modern Industry 4.0 automation settings are characterized by a convergence of IT and OT infrastructure \cite{b15}. This convergence comes with increasingly complex activity patterns and network topologies that make extensive use of autonomous systems and components such as  AI-enabled software applications \cite{b3}. While this has the potential to substantially improve the flexibility, reliability and efficiency of industrial systems  and consumer-oriented manufacturing, it also poses new cybersecurity challenges \cite{b16} and demands a high degree of domain-specific knowledge from analysts assessing potential integrity issues or indications of security compromises. \\
Therefore, there is a clear need for XAI that enables analysts to understand how the system is reaching its conclusions and allow them to interact with it in a collaborative manner \cite{b1}.
One of the biggest drivers for successful adoption of ML models is how well human users can understand and trust their functionality. The benefits afforded by explanations only fully come to bear when these are human-centered  and the users are able to understand and interact with them. This is especially crucial in the cybersecurity domain, where experts require far more information from the model than a simple binary output for their analysis \cite{b3}. \\
A possible example of a cyber attack in an OT system is a security breach. Such a security breach could be a network host that should stay in the local network connecting to the Internet or a developer host directly accessing an edge device. 
In the latter scenario, with a binary output, a connection of the form $192.168.0.80\ to\ 192.168.0.17$ would be flagged as suspicious by the ML model. This gives the analyst very little information about what happened and requires further in-depth analysis. The explanation for the anomaly  {\footnotesize \fontfamily{qcr}\selectfont Security Breach is something whose Connection receives  Service via SSH} given by our approach, shows that edge devices are usually not the origin or destination of SSH connections and why this connection was flagged as suspicious. \\
\\
It is in this context that we propose combining symbolic and sub-symbolic methods on knowledge graphs in order to improve the explainability and quality of IDS-generated alerts in modern industrial systems, increasing their usefulness for analysts. 
This is achieved by integrating domain-specific data, which enables us to better contextualize and enrich cybersecurity-relevant observations, while allowing the sub-symbolic machine learning method GNN to leverage this additional context, i.e., to learn from these observations in a way that makes use of the rich set of interconnections and relations between different entities in the graph.  Similarly to  other  connectionist models, GNNs lack transparency in their decision-making. A variety of explainable models have been developed \cite{b17}, but since such sub-symbolic models are  built for AI researchers, they are often hard to understand for non-experts. We strive to go beyond that by justifying predictions with background knowledge in a human-understandable way by employing a hybrid method \cite{b18}.\\

In this paper, we contribute an explainable methodology for system monitoring in the cybersecurity domain comprised of:
\begin{itemize}
    \item Ontology creation that  follows  the  separation  of  the  industrial automation system into three domains following structured language for cyber threat intelligence (STIX). 
    \item Creating and validating explanations of security alerts through using inductive logic learning and calculating fidelity scores following the approach in \cite{b18}.
    \item Verbalization of the explanations with state-of-the-art verbalization framework \cite{b32} for increased user-friendliness. 
    \item Testing of our proposed system on a hardware demonstrator integrating IT and OT technologies and showing that the proposed method is capable of explaining all triggered alerts which are true positives, and filters out a large percentage of false positives. 
\end{itemize}

To the best of our knowledge, no work exists that leverages XAI to address the alert flooding problem and combines symbolic with sub-symbolic methods for user-friendly alert explanations. 
The rest of the paper is organized as follows. Section II
introduces knowledge graphs and presents a brief overview
of existing applications in the ``XAI for cybersecurity'' domain. In Section
III we describe the anomaly detection, ontology creation and explanation generation method employed
for our specific security monitoring application. Section IV
describes the qualitative and quantitative evaluation of the explanations based on an industrial
automation hardware prototype, before finally drawing some conclusions in Section V.

\section{Background and Related Work}
\subsection{XAI in the Cybersecurity Domain}
XAI in cybersecurity has been gaining more attention in the last two years. The authors of \cite{b1} and \cite{b27} focus on what constitutes a good explanation for the user, including  what information requirements a human
needs for decision-making.  In \cite{b3}, the authors propose a taxonomy for XAI methods and introduce a black box attack for analyzing
the consistency, correctness and confidence properties of
gradient-based XAI methods. In \cite{b28} a rule extraction process is proposed that allows to explain the causes of cyber threats, while in
\cite{b29} a system  is proposed that combines experts’ written rules and dynamic knowledge generated by a decision tree algorithm. 
Similarly, in  \cite{b30} a concrete proposal for an Explainable Intrusion Detection System including a neural network combined with decision trees is presented, together with an empirical evaluation of its prototype implementation.\\
There are several differences between our work and the approaches mentioned above. Despite sharing the general motivation of studying explanations
in the cybersecurity context, our work not only focuses on extracting human-centric explanations  but also ensures the trustworthiness of the explanation through including a sub-symbolic explanation element. Therefore, a fidelity metric can be given that shows how close the explanation is to the actual decision-making process of the ML algorithm.

\subsection{Knowledge Graph Model and Ontology}
\begin{figure}
  \centering
  \includegraphics[width=\linewidth]{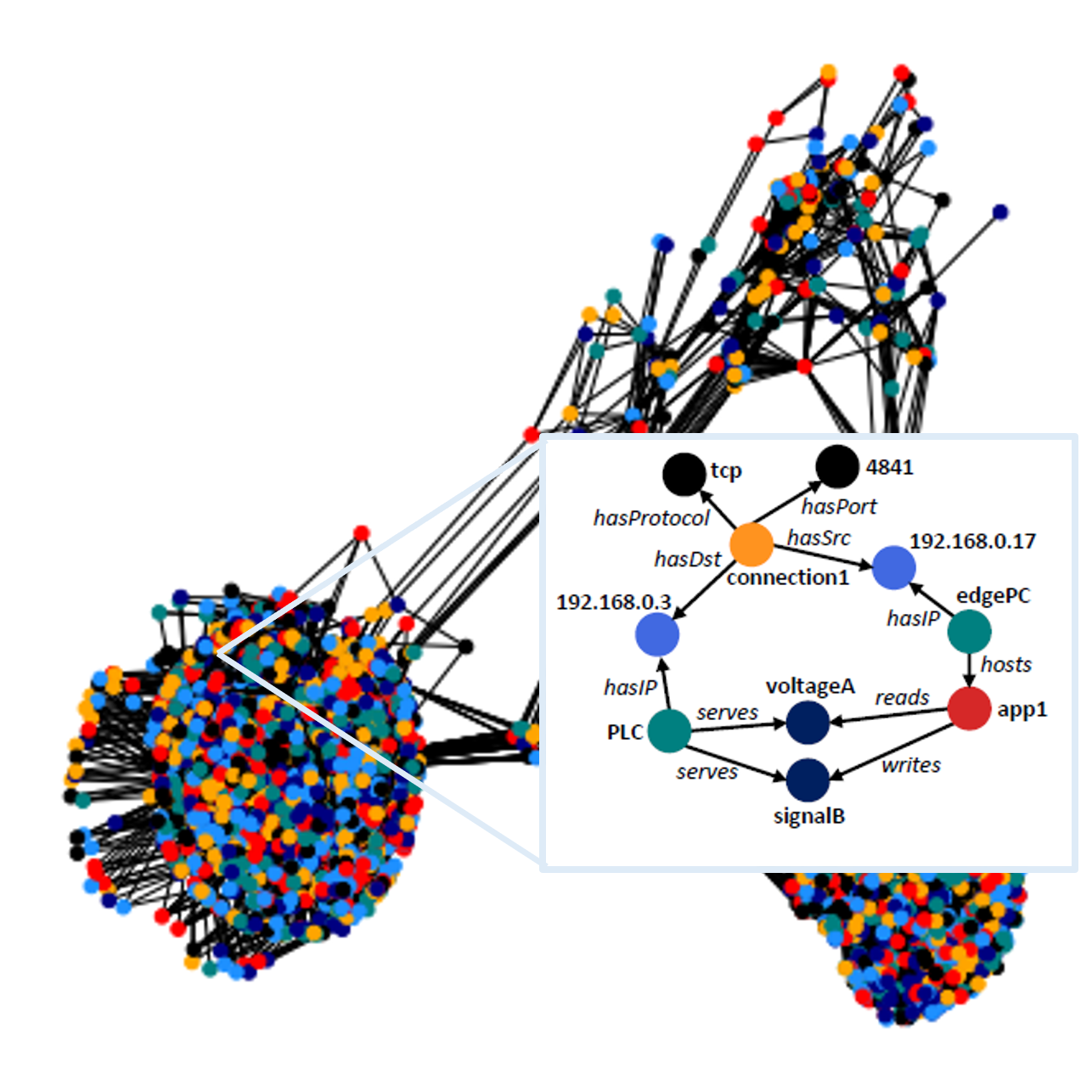}
  \caption{The hardware demonstrator of an OT system is
represented here as a multi-relational graph. Image adapted from \cite{b35} (\copyright 2021 IEEE).}
\label{figure1}
\end{figure}

A knowledge graph is a specific type of knowledge base
where information is encoded in the form of a directed labeled
graph, with nodes representing entities and edges representing
different types of possible relationships between entities.  A simple knowledge graph illustrating these two alternative
representations is depicted in Figure \ref{figure1} for the hardware demonstrator used in this paper. 
Knowledge graphs are particularly useful structures to integrate
data from multiple areas of knowledge, typically making
use of domain-specific vocabularies and ontologies that model
the different categories, relationships, rules and constraints in a specific area of knowledge.  For incorporating explicit domain knowledge into our explanation method on the side of symbolic representation, we use ontologies expressed in the W3C OWL 2 standard\footnote{https://www.w3.org/TR/owl2-overview/} \cite{b26} based on the description logic formalism. The basic constituents for representing knowledge in OWL are individuals, classes and properties. They are used to form axioms, i.e. statements within the target domain, and an ontology
is a set of axioms to describe what
holds true in this domain. The most relevant axioms for our work are class assertions $c(a)$ assigning an individual $a$ to a class $c$, property assertions $r(a_1,a_2)$ connecting two individuals $a_1,a_2$ by property $r$, and subclass axioms $c_1 \sqsubseteq c_2$ expressing that class $c_1$ is a subclass of class $c_2$. Classes can be either atomic class names, such as '{\footnotesize \fontfamily{qcr}\selectfont App1}`, or they can be composed by means of complex class expressions. An example for a complex class expression noted in Manchester syntax is '{\footnotesize \fontfamily{qcr}\selectfont UAVariable and hasDataType byte}`, which refers to all UAVariables which have the data type byte. For details about all types of axioms and the way complex concepts are constructed we refer to \cite{b26}.
\subsection{Graph Neural Networks for Anomaly Detection} 
We apply a GNN to detect unexpected activity in industrial automation systems, which are control systems, such as computers or robots, and information technologies for handling different processes and machineries in a factory. Use of machine learning methods is possible on knowledge graphs, typically by
means of so-called graph embeddings: vector representations
of graph entities which are more suitable for processing
via neural networks and similar methods than their original
symbolic representations.  In \cite{b3}, relational learning on knowledge graphs is applied to security monitoring and intrusion detection by mapping the events in an industrial automation system to links in a knowledge graph.
This way, the anomaly detection task can be rephrased as a link prediction task in the knowledge graph representation of the modeled system.
In particular, \cite{b3} report that the collective learning properties of graph embedding methods allow the resulting models to generalize beyond individual observations, benefiting from the context provided by a rich set of entity and relationship types. Here, we follow the same paradigm of phrasing the security monitoring task as a knowledge graph link prediction task, with a 2-step process. (1)  Learning a baseline of normal behavior by training a GNN on a graph built from
a training dataset. (2) Applying the GNN  in a link prediction 
setting to rank the likelihood of triple statements resulting from events observed at test time and determine whether they represent an anomaly.
A GNN usually consists of graph convolution layers which extract local substructure features for individual nodes and a graph aggregation layer which aggregates node-level features into a graph-level feature vector \cite{b4}. Here, we use a GraphSAGE model \cite{b34}, which is a framework for inductive representation learning on large graphs. It is used to generate low-dimensional vector representations for nodes, and is especially useful for graphs that have rich node attribute information.

\subsection{Explainable AI for Graph Neural Networks}
Current work towards explainable GNNs attempts to convert approaches initially designed for Convolutional Neural Networks (CNNs) into the graph domain \cite{b19}. The drawback of reusing explanation methods previously applied to CNNs are their inability to incorporate graph-specific data such as the edge structure. In order to take node feature information into account \cite{b7}, the model-agnostic approach GNNExplainer finds a subgraph of input data which influence GNNs predictions in the most significant way by maximizing the subgraph's mutual information with the model's prediction. A different type of explainability method tries to integrate ML with symbolic methods. The symbolic methods utilized alongside neural networks are quite agnostic of the underlying algorithms and mainly harness ontologies and knowledge graphs  \cite{b20}. One approach is to map network inputs or neurons to classes of an ontology or entities of a knowledge graph. For example, in \cite{b21}  scene objects within images are mapped to classes of an ontology. Based on the image classification outputted by the neural network, the authors run inductive logic learning on the ontology to create class expressions that act as model-level explanations. An ontology-based approach for human-centric explanation of transfer learning is proposed  by \cite{b22}. While there is some explanatory value to these input-output methods, they fail to give insights into the inner workings of a GNN and cannot identify which type of information was influential in making a prediction. In \cite{b18} this gap is bridged by combining the advantages of both approaches. 

\subsection{Explainer Method: Combining Symbolic and Sub-Symbolic Methods}
We  are  applying a  hybrid  method \cite{b18},  within  which  the  coupling  of  the  sub-symbolic explainer method GNNExplainer \cite{b7} with the symbolic DL-Learner \cite{b33} is used to explain GNN instance-level link predictions.  Firstly,  a  GNN  is  trained  on  and  applied  to  training  and  testing  data  and subsequently the sub-symbolic explainer method GNNExplainer is applied, which outputs explainer subgraphs. Secondly, to create explainer classes for the GNN decision making process, DL-Learner is applied  for  a  specific  predicted  category,  with  positive  and  negative  examples labelled  accordingly. An explainer class is a description that
represents a set of individuals by formally specifying conditions on the individuals' properties. It captures the global behavior of a GNN through investigating what input patterns can lead to a specific prediction.  As the DL-Learner can only process ontologies, the background knowledge is mapped to an ontology as described in Section \ref{ontology}.\\
The pool  of  possible  explainer  classes by the DL-Learner are used to generate instance-level explanations through explainer class entailment.  Explainer class entailment is given when an explainer class applies for a certain alert, given the ontology and  as  can  be  derived  by a standard OWL reasoner.  Or in other words, the explainer class is entailed if the  learned  overall  decision-making  pattern  of  the GNN  applies  to  a  specific  alert.   \\
For increased trustworthiness, the fidelity for each explanation is given. Fidelity is  defined  as  the  measure  of  the  accuracy  of  the  student  model  (DL-Learner)  with  respect  to  the  teacher  model  (GNN).  High  fidelity  is  therefore fundamental whenever a student model is to be claimed to offer a good explanation for a teacher model \cite{b8}. The fidelity is defined as the overlap of the explainer subgraph generated by the sub-symbolic explainer method with the entailed explainer class for a specific instance \cite{b18}.

\section{Cybersecurity Application}
\begin{figure}
  \centering
  \includegraphics[width=\linewidth]{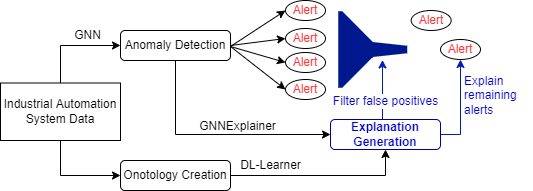}
  \caption{Cybersecurity application workflow.} \label{workflow}
\end{figure}

The high-level workflow for the cybersecurity application which encompasses anomaly detection with a GNN, the creation of an ontology based on the industrial automation system and the explanation generation for the generated alerts can be seen in Figure \ref{workflow}. The individual parts of this workflow are explained in detail below.

\subsection{Experimental Setup}
\begin{figure}
  \centering
  \includegraphics[width=\linewidth]{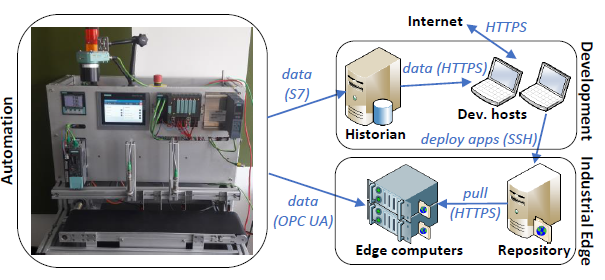}
  \caption{Industrial automation demonstrator consisting of an automation part, a development environment and an industrial edge. Image adapted from \cite{b35} (\copyright 2021 IEEE).} \label{hardware}
\end{figure}

The demonstrator system for e.g. measuring the height of objects for quality control amongst other capabilities, is described in Figure \ref{hardware}, following
the design of modern industrial systems integrating IT and OT
elements. The automation side is equipped with a programmable logic controller (PLC) connected to peripherals via an industrial network.
These include a drive subsystem controlling the motion of
a conveyor belt, an industrial camera, a human-machine interface, and a distributed I/O subsystem with modules
interfacing with various sensors for object positioning and
other measurements (Figure \ref{hardware}, left). The PLC exposes values reported by
these sensors as well as information about the state of the
system by means of an OPC-UA server\footnote{OPC-UA server is a machine to machine communication protocol for industrial automation \cite{b9}.}.
The variables exposed by the server are consumed on
the IT part of the demonstrator by applications hosted on
edge computing servers (Figure \ref{hardware}, bottom right), i.e., computing infrastructure directly
located at the factory floor which is typically devoted to data driven tasks that require close integration with the automation
system and short response times, such as real-time system
monitoring, fault prediction or production optimization.
Industrial edge applications have dynamic life cycles, and
this is captured in the prototype by recreating a development
environment (Figure \ref{hardware}, top right). This cycle starts with development hosts consuming
potentially high volumes of data from a historian, a database that constantly stores process data from the automation
system.  Finally, edge computing hosts fetch
application updates periodically. To make the behavior more
realistic, development hosts occasionally access the internet
with low traffic volumes. The environment is fully virtualized
and performs these activities in an autonomous manner, with
an option to manually induce different types of anomalous
behaviors in order to test the response of our IDS system.
A knowledge graph is built out of the running prototype
by integrating three main sources of knowledge: information
about the automation system, observations at the network
level (e.g., connections between hosts), and observations at the
application level (e.g., data access events). A sizeable portion
of the information is related to the automation system, which is
extracted from engineering tools in the Automation ML format
and ingested into the graph using a readily available ontology
\cite{b10}. Information about application activity is obtained from
the OPC-UA server logs, including session information, i.e., which
variables are accessed and in which way. Finally, all network
traffic is passed through the security monitoring tool Zeek
\cite{b11}, which produces a stream of observed connections that
are ingested using a simple custom data model.

\subsection{Anomaly Detection}
Initially, a baseline is captured with the system under normal
operating conditions, and the collected data is used to train
the link prediction algorithm in an unsupervised
manner. Thereafter, in order to qualitatively evaluate its predictions,
we trigger a set of actions which result in events
not observed during normal operation, but which would be
assigned a wide range of severity levels by a human expert
upon detailed analysis of the available contextual information.
Suspicious behavior is novel behavior given the baseline definition. These scenarios are defined following
the ATT\&CK framework for Industrial Control
Systems\footnote{https://collaborate.mitre.org/attackics/index.
php/Main Page}, i.e., a standardized collection of cyber attack
patterns, to guarantee a high degree of realism. The employed scenarios are listed in Table \ref{scenarios}.
One example is sniffing, where an app accesses data variables completely unrelated to those accessed usually (Scenario 1.2), e.g. not served by the
PLC, or with a different data type, such as strings instead of
numeric data types like int, real, etc. This could be an event
where system information is extracted, like serial
numbers of devices or firmware versions, which
is useful information for discovering back doors and
vulnerabilities of the system. For more details on the scenarios, we refer to \cite{b3}. \\
For the learning we use a 2-layer GraphSAGE model and the Adam optimizer to train the GNN. Training is done for 1000 epochs with learning rate 0.01, reaching a ROC AUC score of at least 84\%. The train/validation/test
split is 80/10/10\%.\\

\vspace{1mm}
\renewcommand{\arraystretch}{1.2}
\begin{table}[ht]
\begin{center}
\begin{tabular}{ |m{29em}|} 
\hline
\textbf{Scenario description} \\
\hline
\textbf{Application activity} \\  
1.1 App changes the way it accesses some variables (e.g. writes instead of reads). \\
\hline
1.2 App accesses variables completely unrelated to those accessed usually.\\
\hline
\textbf{Network activity (HTTPS)}  \\
\hline
2.1 A local address not corresponding to a dev. host (e.g. an edge server) accesses the historian. \\
\hline
2.2 A local address not corresponding to a dev. host (e.g. an edge server) accesses a public IP address. \\
\hline
2.3 A high-volume HTTP access is made to a public IP address (high volumes only from historian in baseline). \\
\hline
\textbf{Network activity (SSH)}  \\
\hline
3.1 The historian host (not a dev. host but on the same network) accesses the app repository via SSH. \\
\hline
3.2 A dev. host accesses an edge server via SSH, but during training no edge servers received SSH connections. \\
\hline
3.3 SSH connection between two edge servers. Usually no edge servers started or received SSH connections.\\
\hline
\textbf{Credential use}  \\
\hline
4.1 Access to OPC-UA server from an IP address that corresponds to a development host. \\
\hline
\textbf{Network Scan}\\
\hline
5.1 Connection which does not match any source-destination pair usually observed. \\
\hline
5.2 Attempt to connect to an IP which is not assigned to any host. \\
\hline
\end{tabular}
\vspace{3mm}
\caption{Attack Scenarios}
\label{scenarios}
\end{center}
\end{table}
\vspace{-8mm}
\subsection{Ontology Creation} \label{ontology}
Analysis of security incidents typically requires consideration
of multiple data sources, some of which are often
exchanged between organizations. In order to facilitate this,
common schemas and data representation formats, such as
STIX \cite{b5}, have been introduced that enable organizations to
exchange threat intelligence in a consistent way. More recently,
these have evolved into fully-fledged ontologies enabling inference
and reasoning \cite{b6}. These ontologies model a wide
range of cybersecurity-relevant knowledge such as product
information, known vulnerabilities and attack patterns, and
can additionally be linked to domain-specific knowledge, e.g.
coming from industrial automation systems \cite{b23}.
Once constructed, these knowledge graphs find a wide range
of applications, e.g., intrusion and threat detection \cite{b24}, \cite{b25}. 
Construction of high-quality knowledge graphs is a challenging
task, especially when it requires extraction of information
from unstructured textual or heterogeneous data.
To use the DL-Learner, we first have to transform the
knowledge graph into an ontology. Here, a class hierarchy that follows the separation of the industrial
automation system into three domains following STIX \cite{b5} is adapted:
\begin{figure}
  \centering
  \includegraphics[width=\linewidth]{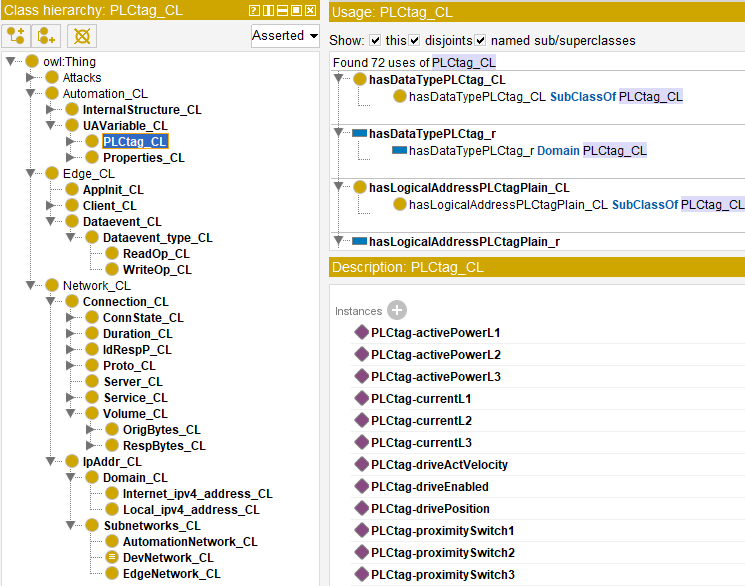}
  \caption{Screenshot of Cybersecurity Ontology in Protégé.} \label{ont}
\end{figure}

\begin{itemize}
\item  \textbf{Automation part}: Summarizes the engineering design of the manufacturing prototype. Further separated into InternalStructure, containing InternalElements, ExternalInterfaces, Attributes and InternalLinks as subclasses as well as UAVariables, containing PLCtags and
Attributes as subclasses.
\item \textbf{Edge part}: Contains app initialization events, data events and the applications.
\item \textbf{Network part}: Contains network connections and their properties as subclasses, as well as IPs and their
subdomains with local and global and automation, development and edge networks as subclasses.
\end{itemize}
For relations, domain and range are provided.
To enable DL-Learner to use properties (like network
connection properties), these have to be promoted to
classes, e.g., every possible instance for network volumes
is its own class. We do the same for applications, InternalElements
and ExternalInterfaces, network types (DevNetwork etc.) and
attributes. Figure \ref{ont} shows a screenshot of the described ontology in the Protégé application \cite{b31}.

\subsection{Explanation Generation}
\begin{figure}
  \centering
  \includegraphics[width=\linewidth]{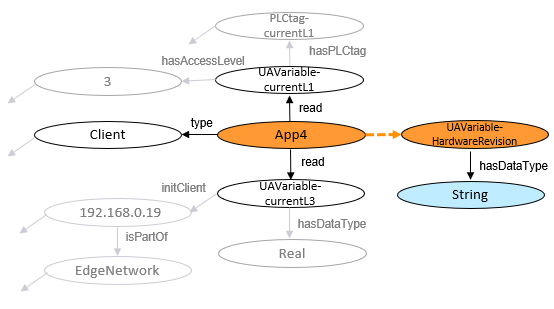}
  \caption{Explanation subgraph for flagging data event "App4 Read UAVariable-HardwareRevision" as suspicious generated by GNNExplainer.} \label{subgraph}
\end{figure}

Tools specially designed to address tasks related to the detection of malicious behavior typically tend to focus more on events or observations that are considered to be unexpected or unusual \cite{b1}. Similarly to that, detected unexpected events serve as the trigger for explanations in our approach, where the type of abnormality is identified.
For the testing of the events from
a certain device (e.g., activity of a certain app, or network
connections between two IPs), the events are compared against each other based on class category.
For example, the suspicious class against the baseline. We do this by using model predictions, based on a ranked list with  suspiciousness scores. Thus, one can  compare, e.g., the top entry with compatible baseline events. This is how positive and negative examples for the DL-Learner are generated, which generates class expressions, such as $(data event\_client$ \textbf{some} $App4$)
\textbf{and} $(data event\_variable$ \textbf{some} $(hasDataType$ \textbf{some} (not ($hasDataType\_REAL$)))).  \\
\\
For increased user-friendliness, the class expressions are verbalized further with the state-of-the-art $LD2NL$ framework \cite{b32}. 
Through the verbalization step, the class expression 
$(data event\_client$ \textbf{some} $App4$)
\textbf{and} $(data event\_variable$ \textbf{some} $(hasDataType$ \textbf{some} (not ($hasDataType\_REAL$)))) of the class $Sniffing$ is translated to {\footnotesize \fontfamily{qcr}\selectfont Sniffing is something whose data event client is App4  and whose data event variable is something that has data type something that is not data type real}.\\
\\
Sub-symbolic explanations in the shape of subgraphs are generated by the GNNExplainer, as can be seen in Figure \ref{subgraph}.
Here you can see the as suspicious flagged data event $App4\ read\ UAVariable-HardwareRevision$. The greyed out nodes and edges are considered not influential in the flagging by the GNN, as opposed to the remaining subgraph. This information is then used to calculate the fidelity score of the entailed explainer class for each flagged event. 
As can be seen in the figure, the entailed explanation is part of the identified subgraph, with the data event client being $App4$ and the $UAVariable-HardwareRevision$ having datatype string. Therefore, the explanation generated by the DL-Learner shows fidelity with respect to the GNN.

\begin{table}[h!]
\begin{center}
\begin{tabular}{ |p{4cm}|p{4cm}|} 
\hline
\vspace{1mm}
\textbf{Explainer Class} & 
\vspace{1mm}
\textbf{Verbalization}  \\
\hline$(data event\_client$ \textbf{some} $App5$)
 \textbf{and} ($data event\_operation$ \textbf{some} WriteOp) 
 
$\mapsto$ Attack Scenario 1.1  & {\scriptsize \fontfamily{qcr}\selectfont  Sniffing is something whose data event client is an App5 and whose data event operation is a Write Operation}  \\
\hline
$(data event\_client$ \textbf{some} $App4$)
\textbf{and} $(data event\_variable$ \textbf{some} $(hasDataType$ \textbf{some} (not ($DataType\_REAL$))))

$\mapsto$ Attack Scenario 1.2  &
 {\scriptsize \fontfamily{qcr}\selectfont Sniffing is something whose data event client is App4  and whose data event variable is something that has data type something that is not data type real}\\

\hline$Network\_CL$ \textbf{and} $(service $ \textbf{some} $ service\_ssh)$  
 
$\mapsto$ Attack Scenario 3.2  &
 {\scriptsize \fontfamily{qcr}\selectfont  Security Breach is something whose service is SSH } \\
\hline
$(id.resp\_p $ \textbf{some} $ id.resp\_p\_22.0)$ \textbf{and} $(orig\_bytes$ \textbf{some} $\ orig\_bytes\_log\_10to2)$  
 
$\mapsto$ Attack Scenario 3.3 &
 {\scriptsize \fontfamily{qcr}\selectfont Security Breach is something whose port is an 22 and whose origin volume is $10\ to\ 2$} \\
\hline 
$(init\_client $ \textbf{some} $  App3)$  \textbf{and} $(init\_server $ \textbf{some} $ (isPartOf $ \textbf{some} $ DevNetwork))$  
 
$\mapsto$ Attack Scenario 4.1  &
 {\scriptsize \fontfamily{qcr}\selectfont 
 Credential Use is something whose initial client is an App3 and whose initial server is 
 something that is part of a development network}\\
\hline
$id.resp\_p$ \textbf{some} $ id.resp\_p\_22.0$

$\mapsto$ Attack Scenario 5.1  &  {\scriptsize \fontfamily{qcr}\selectfont 
Network Scan is something whose port is 22}\\
\hline
\end{tabular}
\vspace{3mm}
\caption{Explainer Classes with Verbalization}
\label{expl}
\end{center}
\vspace{-5mm}
\end{table}

\section{Evaluation}
\subsection{Qualitative Results}\label{qual}
In Table \ref{expl}, a selection of explanations can be seen, along with their verbalization and correspondence with a sub-scenario.  For example, the verbalized explanation  {\footnotesize \fontfamily{qcr}\selectfont Credential Use is something whose initial client is an App3 and whose initial server is  something that is part of a development network} corresponds to Scenario 4.1. "Access to OPC-UA server from an IP address that corresponds to a development host." The explanation captures the anomaly and is even more specific to the concrete data example, as it also gives information about the initial client. This explanation of a Credential Use anomaly is, for example, entailed for the triple $App3\ initiatedFrom\ 192.168.0.80$. \\
The explanation {\footnotesize \fontfamily{qcr}\selectfont Network Scan is something whose port is 22}, which is, amongst others,  entailed for connection  $192.168.0.18\ to\ 192.168.0.60$,  may need some additional information for a layman, but should give ample information for a domain expert. Here, the IP Address does a network scan and, of course, also scans IPs it normally connects to, but with the wrong port - SSH (22) instead of HTTPS (443). Overall we can see that the explanations capture the general scenarios, while often being more specific in describing the concrete anomaly in the data.

\renewcommand{\arraystretch}{1.2}
\begin{table}

\begin{tabular}{ |p{2.1cm}||p{0.9cm}|p{0.8cm}|p{0.8cm}|p{0.8cm}|p{0.8cm}|  }
\hline
 & \textbf{Manual} & \textbf{with Explanation}  & 	reduced by \% & 	\textbf{Fidelity $>$ 0} & 	reduced by \%  \\
 \hline
\textbf{Total (TP + FP)}	 & \textbf{105} & 	\textbf{78} & 	\textbf{0.26} & 	\textbf{42} & 	\textbf{0.60} \\ 

App. activity  & 	12 & 	12 & 	0 & 	6	 & 0.50\\
Network (HTTPS) & 	29 & 	23 & 	0.21 & 	18  & 	0.38\\
Network (SSH)	 & 23 & 	13 & 	0.43 & 	11 & 	0.52\\
Credential Use & 	12	 & 6	 & 0.50	 & 3	 & 0.75\\
Network Scan	 & 29	 & 23	 & 0.21 & 	4	 & 0.86\\
\hline
\textbf{TP (total)}	 & \textbf{64}	 & \textbf{64}	 & \textbf{0}	 & \textbf{39}	 & \textbf{0.39}\\

App. activity	 & 11	 & 11	 & 0	 & 6	 & 0.45\\
Network (HTTPS) & 	23 & 	23	 & 0	 & 18 & 	0.22\\
Network (SSH) & 	8	 & 8	 & 0	 & 8	 & 0\\
Credential Use	 & 3 & 	3 & 	0 & 	3	 & 0\\
Network Scan	 & 18 & 	18	 & 0	 & 4 & 	0.78\\
\hline
\textbf{FP (total)}	 & \textbf{41}	 & \textbf{14}	 & \textbf{0.66}	 & \textbf{3} & 	\textbf{0.93}\\

App. activity & 	1 & 	1	 & 0	 & 0 & 	1\\
Network (HTTPS) &  	6	 & 0 & 	1	 & 0 & 	1\\
Network (SSH)	 & 14 & 	5	 & 0.64 & 	3 & 	0.79\\
Credential Use	 & 9 & 	3	 & 0.67	 & 0 & 	1\\
Network Scan	 & 11 & 	5	 & 0.55	 &  0 & 	1\\
\hline
\end{tabular}

\vspace{3mm}
\caption{Quantitative Results for all as suspicious flagged data events (True Positives TP, False Positives FP)}
\label{quan}
\end{table}

\vspace{-1mm}

\subsection{Quantitative Results}
The training data encompasses more than 37k data events, with 4347 nodes and 37 edge types. The testing data contains 1367 data events, evenly distributed across all 5 attack scenarios, where we reach a ROC AUC score of 84\%. A total of 16 explainer classes have been created. As you can see in Table \ref{quan}, a total of 105 data events have been predicted as unexpected, or in other words, 105 alerts have been triggered, with nearly 40\% being false positives. The learned explainer classes apply to only 14 of the total 41 false positives, while covering 100\% of all true positive alerts. This gives the domain expert the possibility to filter out all triggered alerts that have been created based on the availability of an explanation, reducing the need to investigate false positives by 66\%. No false negatives are created trough filtering, meaning all relevant alerts are still shown to the domain expert. In a further step, the fidelity of the explanation per alert can be taken into account. Such an additional filtering step would lead to a reduction of the false positives by 93\%. However, 39\% of true positives would be missed, which translates to 23\% false negatives in the total alerts. False negatives should be avoided, as critical alerts could be missed. 
Therefore, depending on the preferences of the domain expert, it might be preferable to use the fidelity score as a means to prioritize the alerts. The use of these filter and prioritization techniques significantly reduces the time and resources need by the domain expert.
Additionally, more time will be saved in the analysis of the remaining alerts, as explanations for these are available as seen in Section \ref{qual}. While this is a relatively small example, we can see the potential benefit of applying such a method in a large-scale OT system.

\section{Conclusion}
The  continuous  increase  in  cyber  attacks  has  given  rise  to  a growing  demand  for  modern  intrusion  detection  approaches that leverage ML, e.g. GNNs. However, these methods tend to come with alarm flooding  problems and  a  lack  of  explainability. In this paper,  we  are  addressing XAI for a cybersecurity application by exploring  the combination of  symbolic  and  sub-symbolic  methods that  incorporate  domain  knowledge. We experiment with this approach by generating explanations in an industrial demonstrator system, which are validated through a fidelity score, increasing their trustworthiness. Through empirical and qualitative evaluation, we show that the proposed method produces intuitive explanations for alerts in a diverse range of scenarios. Not only do the explanations provide deeper insights into the alerts, but they also lead to a reduction of false positive alerts by 66\% and therefore address the alarm flooding issue.

\end{document}